\documentclass[aps,prl,showpacs,groupedaddress,twocolumn,preprintnumbers,amsmath,amssymb]{revtex4}

\usepackage{graphicx}% Include figure files
\usepackage{dcolumn}% Align table columns on decimal point
\usepackage{bm}% bold math
\usepackage{braket}

\begin{document}

\title{Non-equilibrium dynamics induced by miscible-immiscible transition in binary Bose-Einstein condensates}
% Force line breaks with \\

\author{Yujiro Eto$^{1}$}
\author{Masahiro Takahashi$^{1}$}
\author{Masaya Kunimi$^{2}$}
\author{Hiroki Saito$^{2}$}
\author{Takuya Hirano$^{1}$}
\affiliation{%
$^{1}$Department of Physics, Gakushuin University, Toshima, Tokyo 171-8588, Japan\\
$^{2}$Department of Engineering Science, University of Electro-Communications, Chofu, Tokyo 182-8585, Japan
}

\date{\today}% It is always \today, today,
             
\begin{abstract}
The non-equilibrium spatial dynamics in a two-component Bose-Einstein condensate were excited by controlled miscible-immiscible transition, 
in which immiscible condensates with domain structures are transferred to the miscible condensates by changing the internal state of $^{87}$Rb atoms.
The subsequent evolution exhibits the oscillation of spatial structures involving component mixing and separation.
We show that the larger total energy of the miscible system results in a higher oscillation frequency. 
This investigation introduces a new technique to control the miscibility and the spatial degrees of freedom in atomic Bose-Einstein condensates.
\end{abstract}

\maketitle

Dynamical pattern formation is one of the most important research subjects in a wide range of fields from non-equilibrium physics \cite{Cross93} to cosmology \cite{Cruz07}.
Bose-Einstein condensates (BECs) in ultracold atomic gases are considered to be a versatile source for such studies by virtue of recent experimental developments, such as the realization of systems with various degrees of miscibility \cite{Kurn98,Modugno02,Eto15PRA2}, the control of atomic interactions \cite{Inouye98,Thalhammer08,Nicklas11,Nicklas14}, and the creation of BECs with strong dipole interactions \cite{Griesmaier05,Lu11,Aikawa12}.
Such unprecedented controllability is conducive to the realization of rich pattern formation dynamics followed by interaction quenching.
They exhibit the $d$-wave collapse in dipolar BEC \cite{,Lahaya08,Aikawa12}, the scaling law in growing patterns \cite{Karl13pra,Karl13scirep,Hofmann14,Nicklas15}, and the Kibble-Zurek mechanism \cite{Sabbatini11,Sabbatini12}.

In experiments using binary systems, a widely used method to study the non-equilibrium pattern formation dynamics is to generate two-component BECs by application of a $\pi/2$ pulse to a single-component BEC.
The component separation and pattern formation dynamics have been widely investigated using this technique \cite{Stenger98, Hall98, Miesner99, Mertes07, Papp08, Anderson09, Tojo10, Egorov11,Egorov13,Eto15,Wang13}.
The miscibility between the two components in these dynamics plays a crucial role, which is determined by the intra- and interspecies $s$-wave scattering lengths. 
The miscibility is therefore tunable by controlling the scattering length between atoms.
Such miscible-immiscible transition in two-component BECs can be realized using magnetic Feshbach resonance \cite{Papp08,Tojo10,Wang13} and dressed states \cite{Lin11,Nicklas15}.

In this Letter, the miscible-immiscible transition in binary BECs are accomplished without the use of the previously mentioned techniques.
We first prepare immiscible BECs and they exhibit component separation. 
The system is then suddenly changed from immiscible to miscible, which imprints the domain structure to the miscible BECs.
The oscillation of spatial structures in the miscible BECs is observed and the frequency of the induced oscillation is found to be closely related to the total energy of the resultant miscible system.
Such highly non-equilibrium dynamics of spatial patterns in miscible BECs have not been studied in a controlled manner to date.

\begin{figure}[b]
\includegraphics[width=8cm]{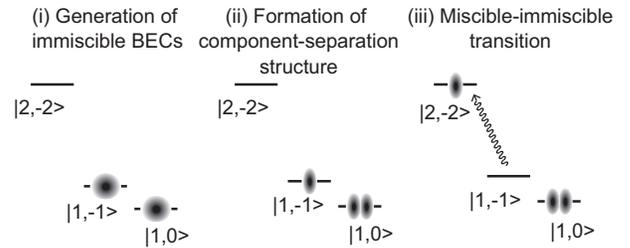}
\caption{
Schematic illustration for generation of the unstable spatial structures in miscible BECs.
The scheme consists of three steps.
(i) Immiscible two-component BECs are generated.
An immiscible pair of $\ket{1,-1}$ and $\ket{1,0}$ is used in this experiment.
(ii) The immiscible BECs spontaneously form the component-separation structures.
(iii) The $\ket{1,-1}$ state is transferred to the $\ket{2,-2}$ state to generate miscible BECs with unstable spatial structures.
}
\label{f:schematic}
\end{figure}

We briefly explain the scheme used to induce the miscible-immiscible transition and excite the non-equilibrium spatial dynamics in miscible BECs.
Firstly, overlapping immiscible BECs comprising $\ket{F = 1, m_F = -1}$ and $\ket{1,0}$ are generated [Fig. 1(i)].
Component-separation forms spatial structures after some evolution of time [Fig. 1(ii)].
The $\ket{1,-1}$ state with a spatial structure is then transferred to the $\ket{2,-2}$ state [Fig. 1(iii)].
The resultant system, which comprises $\ket{2,-2}$ and $\ket{1,0}$, consists of miscible BECs with an unstable spatial structure

The setup used in these experiments is discussed here.
$^{87}$Rb BEC containing $2.9(3)\times10^5$ atoms in the $\ket{ 2, -2}$ state is produced in a crossed far-off-resonant optical dipole trap (FORT) with axial ($z$ direction) and radial frequencies of $\omega_{\rm a} / (2\pi) = 17$ Hz and $\omega_{\rm r} / (2\pi) = 135$ Hz 
(see Ref. \cite{Eto13APEX} for a more detailed description).
To create the stable magnetic field environment,
the entire experimental setup is installed inside a magnetically shielded room.
The bias magnetic field along the $z$ direction is created using a low noise current source (Newport, LDX-3232-100V), and has a magnitude $B_{z} = 11.599$ G.
The immiscible system comprising $\ket{1,-1}$ and $\ket{1,0}$ is generated by the application of a microwave (mw) $\pi$ pulse and radio-frequency (rf) $\pi/2$ pulse, which are resonant to the transition between $\ket{2,-2}$ and $\ket{1,-1}$, and the transition between $\ket{1,-1}$ and $\ket{1,0}$, respectively \cite{Eto15PRA2}.
The mw $\pi$ pulse transfer 99 $\pm 1 \%$ atoms to $\ket{1,-1}$, and the rf $\pi/2$ pulse transfer  50 $\pm 3 \%$ atoms to $\ket{1,0}$.
The immiscible system is converted to a miscible system comprising $\ket{2,-2}$ and $\ket{1,0}$ by application of the mw $\pi$ pulse again.
The intra- and interspecies $s$-wave scattering lengths of the $\ket{1,-1}$ and $\ket{1,0}$ states are $(a_{-1}, a_{0}, a_{-1,0}) = (100.40, 100.86, 101.09)a_{\rm B}$ in the units of the Bohr radius $a_{\rm B}$, and those of the $\ket{2,-2}$ and $\ket{1,0}$ states are $(a_{-2}, a_{0}, a_{-2,0}) = (98.98, 100.86, 97.4)a_{\rm B}$ \cite{Widera06,Kaufman09}, where the subscripts indicate $m_{F}$.
Each set satisfies the immiscible condition ($a_{-1,0} - \sqrt{a_{-1} a_{0}} > 0 $) and the miscible condition ($a_{-2,0} - \sqrt{a_{-2} a_{0}} < 0$).
After the BECs are released from the FORT, each $m_{F}$ component is imaged by the Stern-Gerlach method with a time-of-flight (TOF) of 15 ms.

\begin{figure}[t]
\includegraphics[width=8cm]{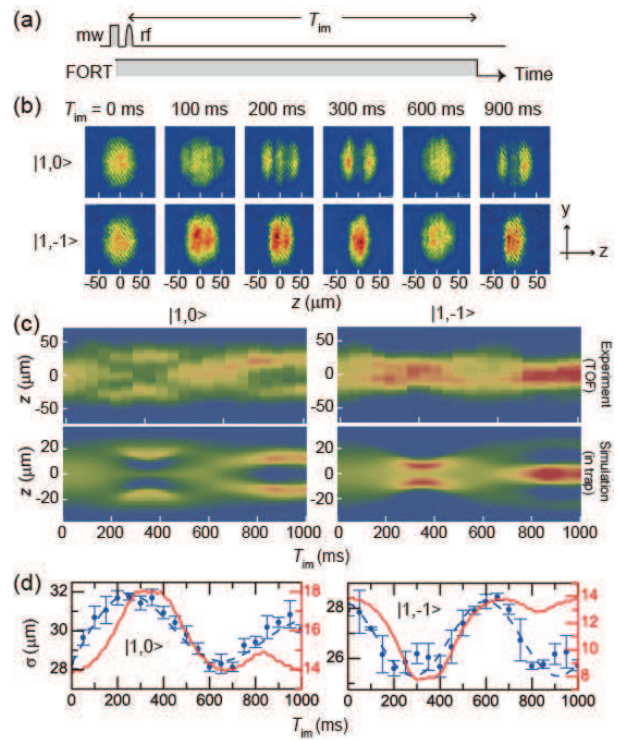}
\caption{
(color online) Pattern formation in immiscible BECs.
(a) Timing diagram for the generation and evolution of immiscible BECs.
The envelopes of mw $\pi$ and rf $\pi/2$ pulses to generate the immiscible BECs are respectively a square wave with a duration of 100 $\mu$s and a Gaussian shaped wave with a standard deviation of 23 $\mu$s.
(b) Typical absorption images of $\ket{1,0}$ and $\ket{1,-1}$ obtained at $T_{\rm im} =$ $0$, $150$, $200$, $300$, $600$, and $900$ ms.
(c) Atomic density distribution with respect to $T_{\rm im}$ integrated over the $y$ direction for $\ket{1,0}$ and $\ket{1,-1}$.
Each distribution corresponds to an average over three measurements. 
The top and bottom panels indicate the experimental data and numerical simulation of in-trap dynamics obtained from coupled Gross-Pitaevskii equations, respectively.
(d) $T_{\rm im}$ versus the spatial width of BECs which is defined as the standard deviation in the $z$ direction.
The circles and dashed curve associated with the left vertical axis represent the experimental data and the damped sinusoidal fit, respectively,
and the solid curve associated with the right vertical axis shows the numerical simulation results.
The error bars are given by the standard deviation over three measurements.
}
\end{figure}

The spatial evolution of overlapping immiscible BECs is investigated using the experimental sequence shown in Fig. 2(a),
where the immiscible BECs are evolved in the FORT during $T_{\rm im}$.
Figure 2(b) shows typical absorption images observed for $T_{\rm im} =$ $0$, $100$, $200$, $300$, $600$, and $900$ ms.
The overlapping immiscible BECs are generated at $T_{\rm im} =$ $0$ ms, 
and then various spatial structures are spontaneously formed.
Figure 2(c) shows the $T_{\rm im}$ dependence of the atomic density distributions integrated over the $y$ direction for $\ket{1,0}$ and $\ket{1,-1}$.
The spatial structure exhibits an oscillatory behavior; after the system evolves toward the component-separated ground state for the first $\sim$ 300 ms,
the initial structure is revived at $T_{\rm im} =$ 600 ms, and then component separation occurs again for another 300 ms ($T_{\rm im} =$ 900 ms). 
This behavior is similar to experiments previously reported in Refs. \cite{Mertes07,Egorov13}.
The standard deviations of the atomic density distributions, $\sigma = (\langle z^2 \rangle - \langle z \rangle^2)^{1/2}$, are calculated to evaluate the frequency of the oscillating structures [Fig. 2(d)].
The oscillation frequency is determined as $\omega_{\rm im}^{\rm fit} /(2\pi) = 1.4(2)$ Hz from the damped sinusoidal fits of experimental data for $\ket{1,0}$ and $\ket{1,-1}$, which is smaller than the axial frequency. 

Although the experimental results obtained after TOF [top panel in Fig. 2(c)] are in reasonable agreement with the in-trap pattern obtained by the simulation (bottom panel),
the fine structures have disappeared due to the effect of TOF and the finite resolution of the imaging system ($\sim$ 7.5 $\mu$m).
In addition, the asymmetry along the $z$ direction appears only in the experimental data, 
which is likely due to the asymmetry of the trap.

\begin{figure*}
\includegraphics[width=18cm]{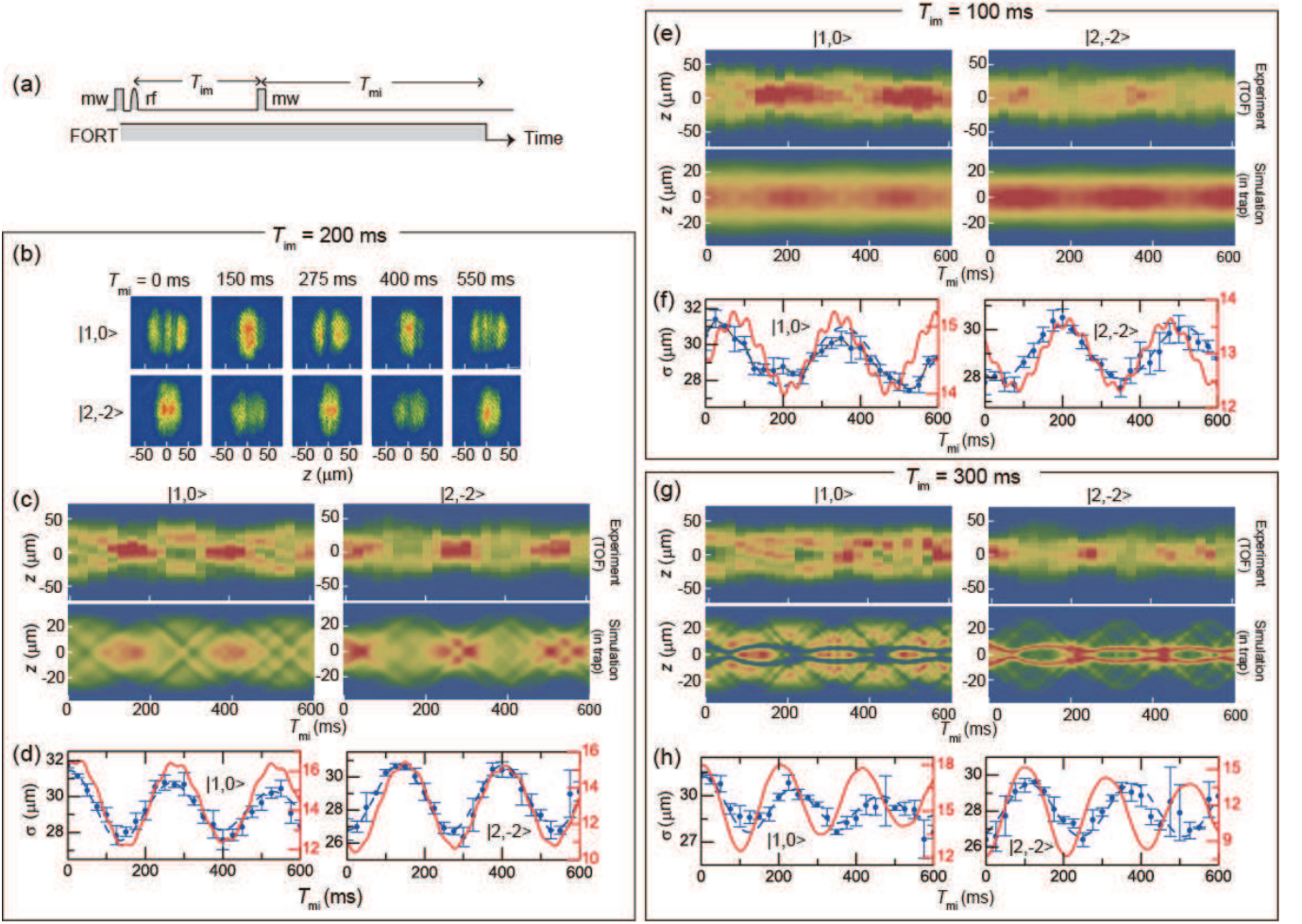}
\caption{(color online) Pattern formation in miscible BECs induced by the transfer imprinting technique.
(a) Timing diagram for the generation of miscible BECs with unstable structures and their evolution. 
$T_{\rm im}$ is set to either $200$ ms [(b)-(d)], $100$ ms [(e) and (f)], or $300$ ms [(g) and (h)].
(b) Typical absorption images of $\ket{1,0}$ and $\ket{2,-2}$ obtained at $T_{\rm mi} =$ $0$, $150$, $275$, $400$, and $550$ ms.
(c), (e), and (g) Atomic density distribution with respect to $T_{\rm mi}$ for $\ket{1,0}$ and $\ket{2,-2}$.
The top and bottom panels show the experimental data and numerical simulation of in-trap dynamics, respectively.
(d), (f), and (h) Standard deviations of the atomic distributions as a function of $T_{\rm mi}$.
The circles, dashed curve (left vertical axis), and the solid curve  (the right vertical axis)  represent the experimental data, the damped sinusoidal fit, and the numerical simulation, respectively.
}
\end{figure*}

The time evolution of the miscible BECs with an unstable spatial structure was investigated using the experimental sequence shown in Fig. 3(a), 
in which the spatial structures transferred from immiscible to miscible BECs can be changed by tuning the duration of $T_{\rm im}$, and the miscible BECs are evolved during $T_{\rm mi}$.
Figure 3(b) shows typical absorption images for $T_{\rm mi} =$ $0$, $150$, $275$, $400$, and $550$ ms,
where $T_{\rm im}$ is fixed at $200$ ms.
The image obtained at $T_{\rm mi} =$ $0$ ms is almost the same as that at $T_{\rm im} =$ 200 ms in Fig. 2(b), 
which indicates that the spatial structure formed in the immiscible BECs is successfully transferred to the miscible BECs.
Although oscillation of the spatial structure is observed, as in the immiscible case shown in Fig. 2, the generated pattern is quite different from the immiscible BECs.
Two types of component-separation structures are formed during one period of oscillation.
The $\ket{2,-2}$ component is surrounded by the $\ket{1,0}$ component at $T_{\rm mi} = 0$ ms, whereas
$\ket{1,0}$ is surrounded by $\ket{2,-2}$ at $T_{\rm mi} = 150$ ms.
  
Figures 3(c) and 3(d) show  the $T_{\rm mi}$ dependence of the atomic density distributions and $\sigma$, respectively.
The oscillation frequency of $3.79(9)$ Hz for the fitted curve is larger than that for the immiscible case.
Furthermore, the oscillation frequency is changed with respect to the transferred structures.
Figures 3(e)-(h) show the spatial evolutions of miscible BECs started from different spatial structures,
where $T_{\rm im}$ is set to $100$ ms [(e) and (f)] and $300$ ms [(g) and (h)].
The oscillation frequencies are estimated to be $\omega_{\rm mi}^{\rm fit}/(2\pi) = 3.12(8)$ Hz and $4.1(3)$ Hz for $T_{\rm im} =$ $100$ and $300$ ms, respectively.

The pattern formation dynamics in miscible BECs was numerically simulated using the sequence given in Fig. 3(a) with various values of $T_{\rm im}$ 
to investigate the relationship between the transferred structure and the induced oscillation of miscible BECs.
Figure 4(a) shows the spectral densities for the temporal variations of $\sigma$ in miscible BECs
($\propto |\int \sigma(T_{\rm mi}) e^{-i \omega_{\rm mi} T_{\rm mi}} dT_{\rm mi}|^2$) for each value of $T_{\rm im}$.
Even for $T_{\rm im} =$ 0 ms, a collective oscillation is induced because the initial density profile is given by the ground state of $\ket{2,-2}$, which is different from that for the mixture of $\ket{1,0}$ and $\ket{2, -2}$.

\begin{figure}
\includegraphics[width=8cm]{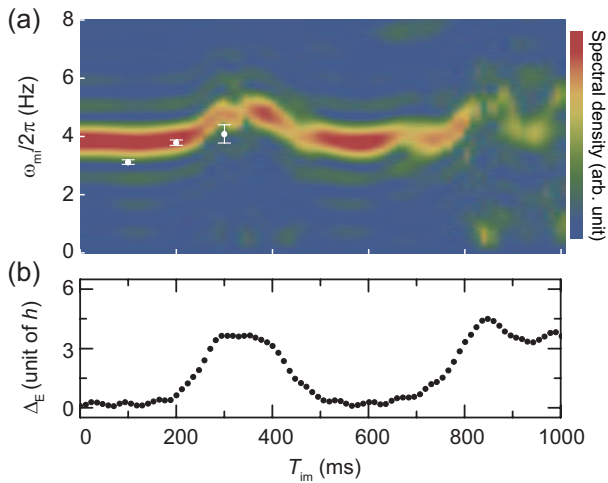}
\caption{(color online) Numerical simulation of the pattern formation dynamics in miscible BECs using the sequence given in Fig. 3(a) for various $T_{\rm im}$.
(a) Spectral density for the temporal variation of spatial width in miscible BECs.
The horizontal axis represents the duration of $T_{\rm im}$.
The spectral densities are calculated from the Fourier transformation of the time series data for $T_{\rm mi} = 0$ ms to $1000$ ms.
The white circles with error bars indicate the experimentally obtained values of $\omega_{\rm mi}^{\rm fit}/2\pi$.
(b) Energy difference $\Delta_{\rm E}$ between the total energy and the ground state energy of the miscible system as a function of $T_{\rm im}$.
 }
\end{figure}

The frequency $\omega_{\rm mi}$ is significantly increased around $T_{\rm im} = 350$ ms, 
at which point the highly separated structures are transferred to miscible BECs. 
This behavior coincides with the total energy of the resultant miscible system, as shown in Fig. 4(b).
The increase in the total energy is caused by the change in the miscibility.
During the process of the component separation in the immiscible system,
the interaction energy changes to kinetic energy. 
When the immiscible-to-miscible transition occurs, the interaction energy increases for the component-separated state with the kinetic energy unchanged.
Thereby the switch of the scattering lengths changes the total energy of the miscible system depending on the spatial structure.
Although the higher energy peak appears around $T_{\rm im} = 900$ ms in Fig. 4(b),
no clear spectral peak is observed in Fig. 4(a) due to the complicated structure oscillations.

In conclusion, we have realized miscible-immiscible transition in two-component BECs by harnessing the rich internal degrees of freedom of $^{87}$Rb atoms.
The non-equilibrium spatial dynamics in miscible BECs is excited by transferring the component-separation structure from immiscible BECs.
The subsequent evolution exhibits structure oscillations that are dependent on the component-separation structures.
According to the numerical simulation, the oscillation frequency is dependent on the total energy of the resultant system.
The numerical result is in good agreement with the experimental result.
The demonstrated technique has various advantages over other techniques;
this technique only relies on the transfer of internal states, 
and external fields such as the magnetic field can be used for other purposes, except at the timing of the transition.
On the other hand, with regard to miscibility control using the magnetic Feshbach resonances and dressed states, the degrees of miscibility are affected by the inhomogeneity of the magnetic fields and the collisional frequency shift, respectively.
In addition, the spin-exchange inelastic loss can be avoided with this technique by selecting appropriate pairs of internal states.
This simple technique could be widely applied to various experimental conditions, e.g., other atomic species.
Such an investigation on the control of miscibility and the spatial degrees of freedom has provided a new technique to examine the non-equilibrium dynamics of multicomponent BECs.

The authors thank M. Horikoshi for fruitful discussions and M. Sadgrove for valuable comments on the manuscript.
This work was supported by Grants-in-Aid for Scientific Research (C) (No. 26400414, No. 15K05233), and a Grant-in-Aid for Scientific Research on Innovation Areas Fluctuation \& Structure (No. 25103007) from 
the Ministry of Education, Culture, Sports, Science, and Technology (MEXT) of Japan, and by a  Research Grant from the Yoshishige Abe Memorial Fund.

\end{document}